\documentclass[review,number,sort&compress]{elsarticle}

\usepackage{amssymb}
\usepackage{epsfig}
\usepackage{graphics}
\usepackage{graphicx}
\usepackage{subfigure}
\usepackage{exscale}
\usepackage{latexsym}
\usepackage{makeidx}
\usepackage[section]{placeins}
\usepackage{xtab}
\usepackage{epsf}
\usepackage{bbm}

\usepackage{lineno}
\journal{Nuclear Instruments and Methods in Physics Research Section A}

\begin{document}
\begin{frontmatter}

\title{Calibration of the compact neutron spectrometer at PTB}

\author{M.~Osipenko$^a$,
V.~Ceriale$^b$,
G.~Gariano$^a$,
M.~Girolami$^c$,
S.~Cerchi$^a$,
R.~Nolte$^d$
E.~Pirovano$^d$
M.~Ripani$^{a,e}$,
D.M.~Trucchi$^c$.
}

\address{
$^a$ \it\small INFN, sezione di Genova, Genova, 16146 Italy, \\
$^b$ \it\small Dipartimento di Fisica dell'Universit\`a di Genova, Genova, 16146 Italy, \\
$^c$ \it\small CNR-ISM, Monterotondo Scalo, 00015 Italy, \\
$^d$ \it\small Physikalisch-Technische Bundesanstalt (PTB), Braunschweig, D-38116 Germany, \\
$^e$ \it\small Centro Fermi - Museo Storico della Fisica e Centro Studi e Ricerche "Enrico Fermi', Rome, 00184 Italy.
}

\begin{abstract}
In the present article we describe the calibration measurements of the compact
neutron spectrometer at certified PTB neutron source.
The spectrometer is based on two diamond detectors enclosing the $^6$Li neutron converter.
The incident neutron energy is measured through the sum of energies of the
two conversion reaction products: $t$ and $\alpha$.
The prototype used in this work was based on two 4$\times$4 mm$^2$, 100 $\mu$m thick
single crystal CVD commercial diamond plates.
% produced by Applied Diamond Inc.
%%%
The measurements were performed in the fission spectrum domain
at five different neutron energies varying from 0.3 up to 2 MeV.
The obtained data showed a very good
neutron energy reconstruction, i.e. in agreement with its reference values within
statistical uncertainties or with deviation below 11 keV in absolute scale.
The energy resolution of the spectrometer was found to be about 100 keV (RMS)
with a small (about 9 keV/MeV) rising trend towards higher energies.
The measured detection efficiency was found to be compatible with expectations
based on analytical calculations and on Geant4 simulations
within 3\% statistical and 4\% systematic precision.
\end{abstract}


\end{frontmatter}

\section{Introduction}\label{sec:intro}

%\linenumbers

The operation of solid state detector in an environment with high neutron flux requires
significant research and development effort. In particular, within the core
of a fission reactor the detector has to be very compact, resistant to large
doses of fast neutrons, fast (for on-line detectors) and largely insensitive to $\gamma$ background.
Among solid state sensors the best choice for such application
is the diamond detector. With respect to the standard Silicon detector the diamond
is more radiation hard~\cite{cvd_rad_hard}, has a lower sensitivity to $\gamma$ background
and can be operated at high temperatures.
With respect to $^3$He and CH$_4$ proportional counters,
usually employed at reactor facilities,
the solid state detector can cover broader energy range maintaining
its compact size and practicable rates.

The spectrometer based on a sandwich of two diamonds was developed in Ref.~\cite{sdw_calib}
to measure neutron energy independently for every neutron interaction event through the sum
of energies deposited by conversion reaction products.
Few prototypes of the spectrometer were tested at fast fission reactor in Ref.~\cite{sdw_tapiro},
in thermal reactor in Ref.~\cite{sdw_tapiro}, at DD and DT fusion neutron source
in Refs.~\cite{sdw_calib,sdw_dt}.
All these measurements demonstrated feasibility of the technique
and emphasized numerous issues which were successfully resolved.
However, all these measurements had a very limited systematic precision
because of relatively poor knowledge of neutron source flux (to the level of 20\%).
Moreover, monochromatic sources were limited to only two neutron energies:
thermal and 2.5 MeV from DD fusion generator. Thus, a systematic
study with many different neutron energies in the region of interest and having flux uncertainties
below 5\% (maximum $^6$Li cross section uncertainty) were necessary to conclude this R\&D project.

In this article we describe dedicated calibration measurements at certified PTB neutron source
performed within FP7 CHANDA project Transnational Access workpackage.
The main goal of this experiment was to demonstrate high precision
of incident neutron energy reconstruction and reliability
of detection efficiency evaluation through Geant4 simulations.
The article is structured in two main parts:
in section~\ref{sec:det} we listed briefly the improvements
of the last prototype of diamond spectrometer operated at PTB,
while in section~\ref{sec:ptb} described the experimental setup used at PTB
and the results of the measurements.

%%%%%%%%%%%%%%%%%DETECTOR%%%%%%%%%%%%%%%%%%%%%%%%%%%%%%%%%%%%%
\section{Detector Description}\label{sec:det}
The detailed description of the spectrometer as well as of the techniques
used to build it was given in Refs.\cite{sdw_upg,sdw_dt,sdw_calib,sdw_tapiro}.
Hence here we describe only few aspects which differ the present prototype
from those already published.
First of all in the present spectrometer
single crystal CVD diamonds from Applied Diamond~\cite{applied_diamond} were used.
We selected 100 $\mu$m thick diamonds as the optimum size to measure
conversion products of neutrons up to about 7 MeV.
The surface area of $4\times 4$ mm$^2$ was covered by thin Au contacts
leaving 200 $\mu$m margins at the borders.
On the inner side of the sandwich contacts were 30 nm thick to minimize energy loss by neutron conversion products,
while on the outer surface thickness was increased to 100 nm.
The deposition procedure developed at ISM-CNR for almost ohmic contacts
described in Refs.\cite{sdw_upg,Trucchi1,Trucchi2,Trucchi3,Trucchi4} was used.

The contacts consist of a multilayer structure composed by a very thin ($<$3 nm) DLC layer,
which was demonstrated to improve ohmicity on intrinsic diamond
and stability of the contact even under high-flux neutron irradiation, and an Au cap layer.
DLC layer was formed by bombardment with keV energy Ar ions,
able to induce amorphization of the diamond surface,
whereas the Au cap layer was deposited by RF magnetron sputtering.
A particular care was dedicated to the control and uniformity of the thickness
of the deposited Au cap layer (30$\pm$10 nm on the inner surface, 100$\pm$20 nm on the outer one);
for this purpose, a calibration chart was drawn by depositing several Au films
with different deposition times on different samples, and then measuring
the corresponding thickness of the deposited films by mechanical profilometer.
%The thickness percentage uncertainty increases at low values
%owing to the increasing uncertainty owing to shorter deposition times.

On the inner surface of one diamond, covered by Au contact, a 200 nm thick $^6$LiF film was thermally evaporated.
LiF film had area of 3$\times$3.2 mm$^2$ in order to leave margins of active diamond surface
and simultaneously to allow electrical connection of the inner contacts to the ground.

\begin{figure}[h]
\begin{center}
\includegraphics[bb=1cm 0cm 20cm 26cm, angle=270, scale=0.3]{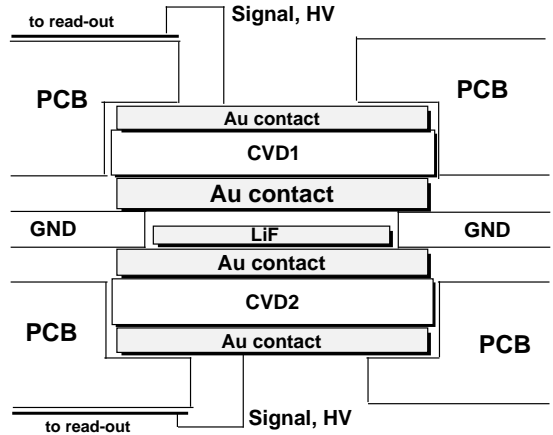}
\caption{\label{fig:ptb_sdw_geom}The drawing (not in scale)
of the spectrometer structure around diamond sensors. All components,
like two PCBs, ground plate (GND), two diamonds (CVD), their Au contacts and LiF converter
are clearly indicated.}
\end{center}
\end{figure}

In the previous prototype from Ref.~\cite{sdw_upg} we suspected that the conductive glue,
used to hold the diamonds, could be the origin of the excessive energy loss by $\alpha$s.
Thus, in the present prototype only mechanical pressure was used to provide
electrical connections of inner contacts to the ground.
This was performed by means of two Nylon screws pressing two PCBs towards each other.
The two diamonds were located in shallow grooves on the two opposite PCBs
while pressing gold-plated 50 $\mu$m thick ground plate in the middle
as shown in Fig.~\ref{fig:ptb_sdw_geom}.
The products of neutron conversion on $^6$Li were subject to energy loss only
in a fraction of LiF film (100 nm in average), 30 nm Au contact and 50 $\mu$m of air
for one of two products.
The outer contacts of the two diamond sensors were connected
to signal readout traces on PCBs (providing also high voltage bias) by wedge bonding
through holes in the PCBs.

In order to improve Signal-to-Noise ratio we implemented the passive
signal preamplification scheme by means of 150 MHz RF transformer developed in Ref.~\cite{sdw_upg}.
This scheme is particularly useful when no Si-based amplifiers could be
place near to the detector due to high flux of fast neutrons.

The outer case of the spectrometer, cables, amplifiers and Data AcQuisition (DAQ) system
were the same as previously described in Ref.~\cite{sdw_upg}.
Important to notice that the first amplifiers were located at a distance
of 3.5~m from the spectrometer where neutron flux was significantly lower.
All the connections were well shielded from electromagnetic interference
by additional wire braid
and isolated from the experimental Hall metal floor.

%%%%%%%%%%%%%%%%%CALIBRATIONS%%%%%%%%%%%%%%%%%%%%%%%%%%%%%%%%%%%%%
\section{Measurements at PTB}\label{sec:ptb}
The spectrometer was calibrated both in the reconstructed neutron energy
and absolute efficiency in the certified neutron fluxes at PTB~\cite{ptb_n}.
Quasi-monochromatic neutron beams in the energy range 0.3-2 MeV were obtained
through proton reactions on thin LiF (0.145 mg/cm$^2$) and TiT (0.529 mg/cm$^2$ Ti at 1.25 MeV
and 0.955 mg/cm$^2$ at 2 MeV) targets. 
The absolute neutron yield was limited by the incident proton beam current
and target thickness, optimized for the best neutron energy resolution.
The spectrometer, designed for reactor measurements, featured an active area
of $4\times 4$ mm$^2$ provided a small geometrical acceptance.
In order to maximize neutron flux incident on spectrometer active surface
it was installed along the proton beam direction (zero degrees)
at the distance of 5.05$\pm$0.05 cm from the neutron production target.
The distance was limited from the lower side by the beam pipe extension
and solid angle covered by detected neutrons, which was equal to 5.85 mstr,
close to that of flux monitoring system.
Since neutron energy was monochromatic only at a fixed polar angle, a larger
solid angle would affect neutron energy resolution.
Moreover, the relative uncertainty on the source-to-spectrometer distance
at 5 cm was nearly equal to the uncertainty on neutron flux calibrations.
In this position the obtained neutron flux varied from 0.15 up to 0.9 $\times$10$^6$ n/cm$^2$/s
depending on beam energy and production target used.
In order to minimize neutron backscattering from surrounding materials
the spectrometer was mounted on 30 cm long and 16 cm high, ``L''-shaped,
lightweight (3 mm thick) Aluminum support. No other material was located
within 30 cm from the spectrometer. The support was attached to the standard
PTB Aluminum stand, whose top end was lowered 20 cm below beamline axis.

During the measurements neutron flux was monitored by long counter
installed at 16 degrees. We also saved into the data stream accumulated
beam charge scaler in order to keep track of time dependent variations of neutron flux.

The data were taken at five different beam energy-target combinations
listed in the Table~\ref{tab:beam_target}. The incident neutron energy
spread $\Delta E_n$ (FWHM) was estimated from a combination of proton energy loss
in the target and angular acceptance of the spectrometer.
The statistics was limited by the granted beam time.

For each neutron energy, the neutron flux monitors were calibrated in a dedicated run.
With the spectrometer placed outside the neutron field,
a De Pangher long counter (PLC) was first positioned at zero degree
and a distance of about 4 m from the neutron production target to measure the neutron fluence.
The shadow cone technique was employed to subtract the contribution of room-return neutrons
to the number of events detected by the PLC~\cite{Nolte_1,Roberts_2}. The contribution of neutrons
scattered from the PLC into the the neutron flux monitor at 16 degrees, during the fluence measurement
was determined in two subsequent short runs with and without the PLC in place.
For these short runs the integrated target current was used for normalization.
After correcting for neutron scattering in air and for the contribution of neutrons
of lower energy produced by scattering of primary neutrons in the target assembly (see below),
this procedure allowed to determine the neutron yield per monitor count,
i.e. the number of neutrons emitted per unit solid angle at a neutron emission angle of 0 degrees
without further interaction with the target assembly.
This number was used during the runs with the spectrometer to obtain the fluence
with an uncertainty of the order of few percent~\cite{Schlegel_3,Schlegel_4}.

\begin{table}[!h]
\begin{center}\label{tab:beam_target}
\caption{Beam-target configurations of different runs:
proton beam energy $E_p$ and current $I_p$,
neutron energy $E_n$, its spread $\Delta E_n$ (FWHM)
and flux at the detector face.} \vspace{2mm}
\begin{tabular}{|c|c|c|c|c|c|} \hline
 $E_p$ &  $I_p$   & Target	      &  $E_n$  & $\Delta E_n$& $\phi_n$	   \\
$[$MeV]& [$\mu$A] &		      & [MeV]   & [keV]       & [n/cm$^2$/s]	   \\ \hline
 2.067 &    9.5   & LiF  	      & 0.301   & 20	      & 0.15$\times$10$^6$ \\ \hline
 2.334 &   10	  & LiF  	      & 0.601   & 16	      & 0.9$\times$10$^6$  \\ \hline
 2.620 &   10	  & LiF  	      & 0.905   & 14	      & 0.4$\times$10$^6$  \\ \hline
 2.094 &    4	  & TiT 0.5 mg/cm$^2$ & 1.2497  & 54	      & 0.4$\times$10$^6$  \\ \hline
 2.821 &    3.5   & TiT 1 mg/cm$^2$   & 1.9986  & 77	      & 0.8$\times$10$^6$  \\ \hline
\end{tabular}
\end{center}
\end{table}

For energy calibration purpose we acquired also three thermalized neutron runs,
measured at the beginning, in the middle and at the end of experiment
with 2, 0.6 and 1.25 MeV reference neutron energy, respectively.
To this end, the spectrometer was attached in front of a 5 cm thick plastic moderator
in such a way that only backscattered neutrons could be detected.
Thanks to steep rise of $n(^6Li,t)\alpha$ cross section at low energy of incident
neutrons the overall event rate of these runs increased by a factor of 20
with respect to runs without moderator
and it was dominated by quasi-thermal neutron interactions.

\begin{figure}[!ht]
\begin{center}
\includegraphics[bb=3cm 0cm 20cm 26cm, scale=0.35, angle=270]{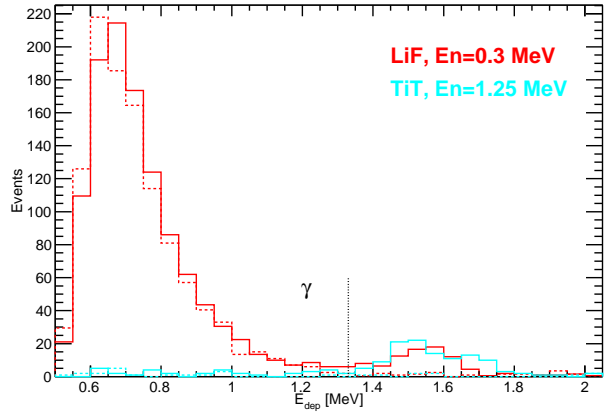}
\caption{\label{fig:ptb_edep12_300vs1250}Energy deposited in the upstream (solid line)
and downstream (dashed line)
diamond sensor of the spectrometer for different neutron production targets
LiF (red) and TiT (cyan). Beam energies were also different as indicated on the plot.
For LiF target the tail rising towards lower energies was due to $\gamma$ background.
Dotted line at 1.3 MeV indicate the maximum energy which can be deposited in coincidence
by Compton electrons in two 100 $\mu$m thick 4 mm wide diamonds.}
\end{center}
\end{figure}

%%% Event Selection
The Data AcQuisition system (DAQ) described in detail in Refs.\cite{sdw_upg}
had the trigger set on the coincidence of signals
from the two diamond sensors above 0.5 MeV threshold within 64 ns gate window.
One run at 1.25 MeV was taken with single sensor trigger and demonstrated
good efficiency of coincidence selection.
The choice of the threshold value was aimed to optimize efficiency for
$n(^6Li,t)\alpha$ events, while reducing contaminations from electronics noise
and $\gamma$ background. In fact, considering the active thickness
of diamond sensor of 100 $\mu$m the expected energy loss
of
% 1 MeV 
Compton electrons crossing its volume along the normal to spectrometer plane (and along the beam)
was about $57$ keV. Therefore, only electrons going at $>$83.5 degrees with respect to
the normal to diamond plane could leave $>$0.5 MeV deposited energy.
However, the lateral size of the sensors combined with the distance between them
allow only $<$87.5 degrees straight tracks to cross both diamonds completely.
This leaves a small acceptance region (of about 0.4 str) for measurable $\gamma$ background
(of the order of 3\% of all Compton conversions in 100 nm Au contact and of nearest 100 $\mu$m of PCB).
Nevertheless, runs measured with LiF target (0.3, 0.6 and 0.9 MeV)
showed significant $\gamma$ contamination at total deposited energies below 2.5 MeV.
The first indication of $\gamma$-background was the similarity of
the low energy tail distributions of deposited energies in two diamond sensors measured with LiF target,
shown in Fig.~\ref{fig:ptb_edep12_300vs1250}. Moreover, from the maximum
angular acceptance for Compton electrons crossing both diamonds mentioned above
we estimated the maximum energy deposited by them in each diamond to be below 1.33 MeV.
Indeed, the low energy peaks in Fig.~\ref{fig:ptb_edep12_300vs1250} vanish at about this energy.
This background
was absent in the runs with TiT target. All above confirmed that the background was due to the
$\gamma$ production on $^{19}$F, perhaps through the high-$Q$ value $p(^{19}F,^{20}Ne^*)\gamma$ ($Q$=12.8 MeV)
or $p(^{19}F,^{16}O^*)\alpha$ ($Q$=8.1 MeV) reactions.
The $\gamma$s with energy $>$3 MeV produced Compton electrons of $>$2.5 MeV at about 85 degrees,
which crossed both diamonds leaving symmetrical signals.
This background was completely removed in the off-line analysis
by applying a software threshold of 2.5 MeV on the total deposited energy,
still well below $n(^6Li,t)\alpha$ reaction $Q$-value.

%

\begin{figure}[!ht]
\begin{center}
\includegraphics[bb=3cm 0cm 20cm 26cm, scale=0.35, angle=270]{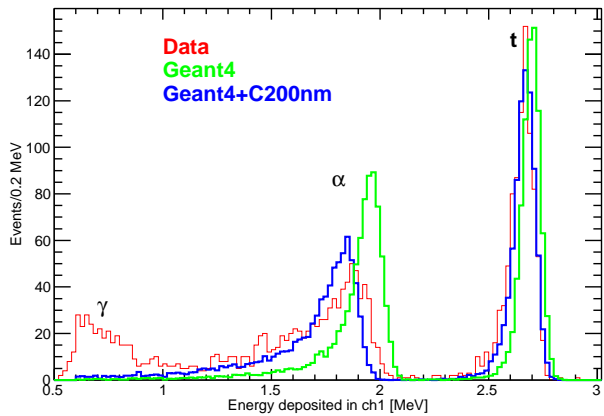}
\caption{\label{fig:ptb_edep_i_th}Comparison of spectrum measured
in single diamond sensor of the spectrometer under quasi-thermal neutron irradiation
with Geant4 simulations.
%The data are corrected for the effective energy loss
% of $t$ 
%by applying an overall energy scale factor of 1+35 keV/2.7 MeV.
Geant4 simulations do not contain $\gamma$-background, seen in the data on LiF target.
}
\end{center}
\end{figure}

%%% Energy Calibration
The energy deposited in each single diamond of the spectrometer was calibrated independently
using the data acquired in quasi-thermal neutron runs. First of all zero energy reference
was extracted from the measured baseline level. To this end in every event
a number of pre-trigger samples exceeding typical signal width was saved.
We checked that the baseline was stable during the whole experiment.
Then spectra of individual diamonds were compared to Geant4 Monte Carlo simulations
as shown in Fig.~\ref{fig:ptb_edep_i_th}.
Because $t$ experienced much lower energy loss in LiF layer and Au contacts
with respect to the $\alpha$, as one could conclude from a comparison of the widths
of the corresponding peaks at 2.7 and 2 MeV, and from Geant4 output, we used
the peak at 2.7 MeV as the second reference to determine the energy scale.
Indeed, $t$ peak RMS in Fig.~\ref{fig:ptb_edep_i_th} was about 50 keV
composed of electronics nose, estimated to be 35 keV,
and about 35 keV related to the energy loss fluctuations.
Instead, the $\alpha$ peak at 2 MeV
exhibited a very asymmetric shape related to large energy losses.
These losses were larger than those expected for 30 nm Au and 100 nm LiF
layers and similar to those observed in Ref.~\cite{sdw_upg}.

\begin{figure}[!ht]
\begin{center}
\includegraphics[bb=3cm 0cm 20cm 26cm, scale=0.35, angle=270]{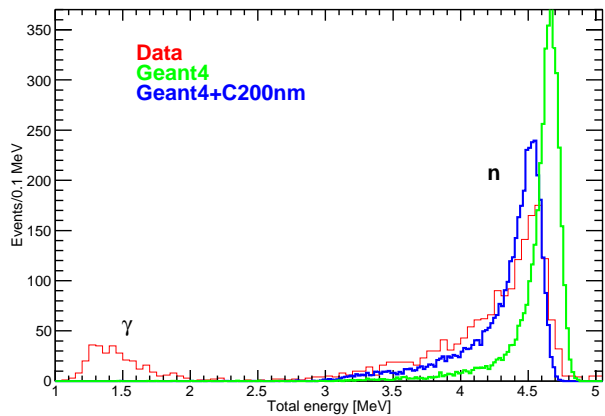}
\caption{\label{fig:ptb_edep_tot_th}The same as in Fig.~\ref{fig:ptb_edep_i_th} but for the
total energy deposited in the spectrometer.}
\end{center}
\end{figure}

\begin{figure}[!ht]
\begin{center}
\includegraphics[bb=3cm 0cm 20cm 26cm, scale=0.35, angle=270]{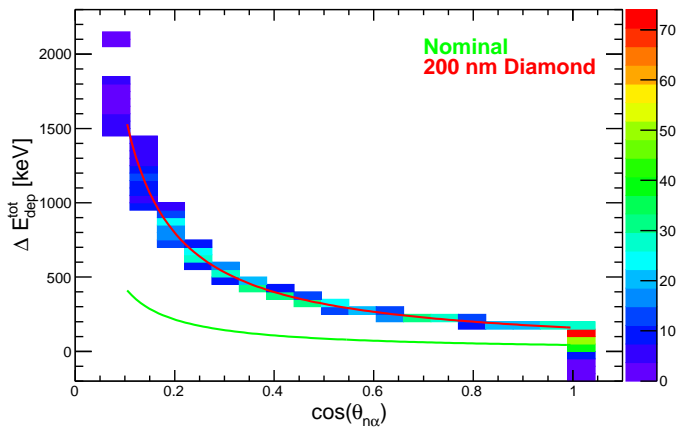}
\caption{\label{fig:ptb_edep_cos}The correlation between the total energy loss by $\alpha$ and $t$
and reconstructed cosine of $\alpha$ production angle with respect to the normal to diamond surface.
The solid lines show analytic calculations assuming nominal materials (green)
and those adding 200 nm dead layer in diamond (red).}
\end{center}
\end{figure}

In order to confirm that the observed excessive energy loss by $\alpha$s
was due to an additional dead layer in diamond we studied their angular distribution.
Using the combination of known neutron energy, beam direction and energies of two conversion reaction
products we reconstructed the $\alpha$ production angle with respect to the beam direction,
which coincided with the normal to diamond surface.
The correlation between the observed energy loss and cosine of $\alpha$ production angle
is shown in Fig.~\ref{fig:ptb_edep_cos}.
The data exhibit a clear $1/\cos{\theta_{\alpha n}}$ behavior, however
the size of the effect is 3.7 times larger than expected using the nominal
thicknesses of LiF and Au contacts. Given the good agreement of the measured
detection efficiency with its expectations we exclude possibility for LiF layer
being thicker. Au contact thickness was controlled during the deposition
with uncertainty of 20 nm, largely insufficient to explain the difference (would require 180 nm).
The only plausible explanation of these data
is the presence of a 200 nm thick dead layer on the surface of both diamonds.
This dead layer was probably formed either during crystal polishing
or during contact fabrication procedure.

%

\begin{figure}[!ht]
\begin{center}
\includegraphics[bb=3.5cm 2cm 20.8cm 25cm, scale=0.35, angle=270]{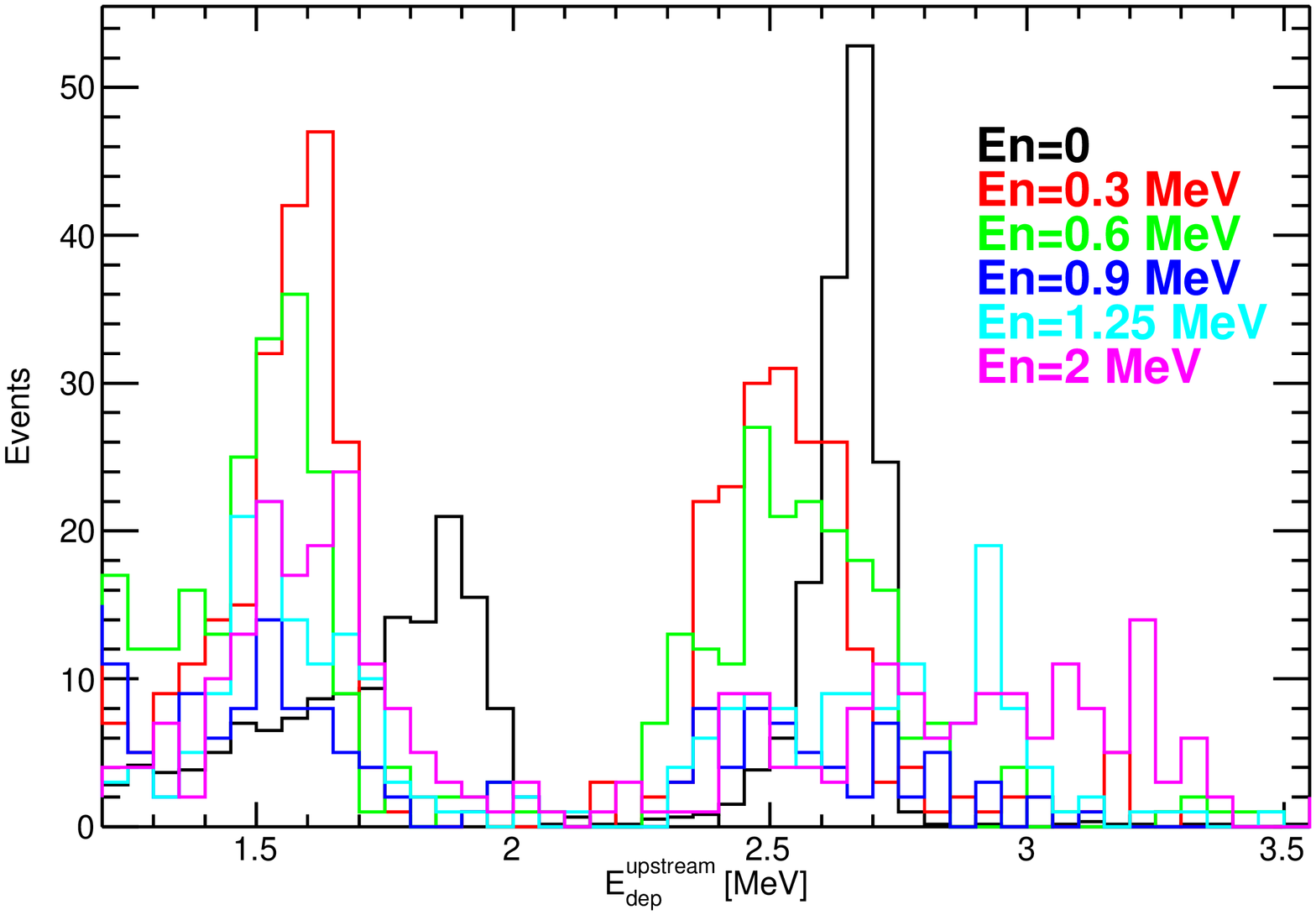}\\
\includegraphics[bb=3.5cm 2cm 20.8cm 25cm, scale=0.35, angle=270]{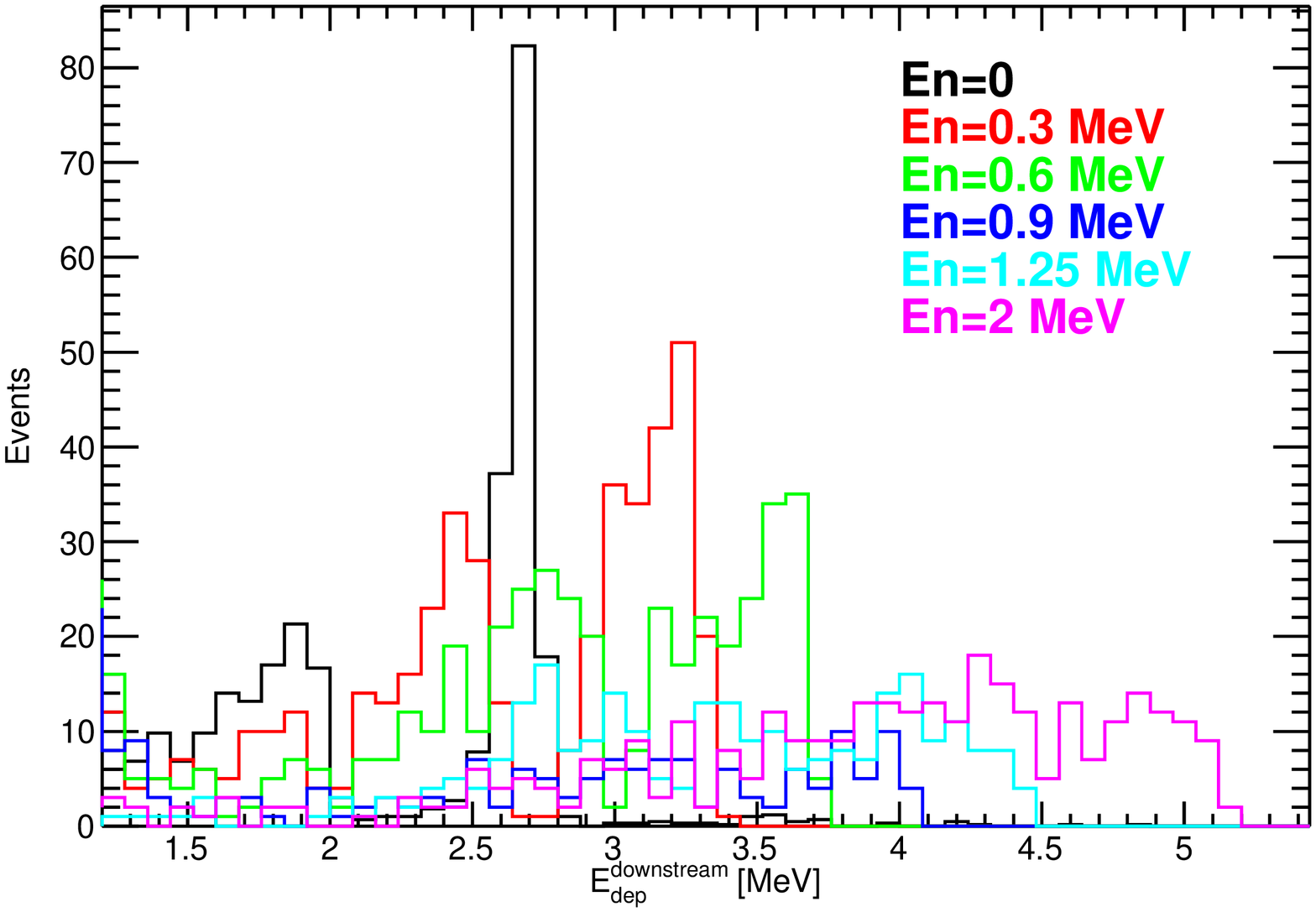}
\caption{\label{fig:ptb_edep12_en}Energy deposited in the upstream (top) and downstream (bottom)
diamond sensors of the spectrometer for different incident neutron energies.}
\end{center}
\end{figure}

%%% Single Diamond Energy Distributions
After the energy calibration of the spectrometer was fixed through the analysis
of quasi-thermal neutron irradiation data we reconstructed deposited energy spectra
for all other beam energies, shown in Fig.~\ref{fig:ptb_edep12_en}.
The distributions in the upstream diamond preserve almost the same shape
with increasing neutron energy because it is largely compensated by the Lorentz
boost of the backward going product. Indeed, $t$ and $\alpha$ peaks remain separated
up to 2 MeV, although the peaks become broader due to product angular distribution.
Instead, in the downstream diamond the energy increase adds directly to the
product energy, broadening it through the angular distribution. In this case
for neutron energies above 0.3 MeV $t$ and $\alpha$ cannot be distinguished anymore
by studying deposited energy spectra.

%

\begin{figure}[!h]
\begin{center}
\includegraphics[bb=3.5cm 0cm 20cm 26cm, scale=0.35, angle=270]{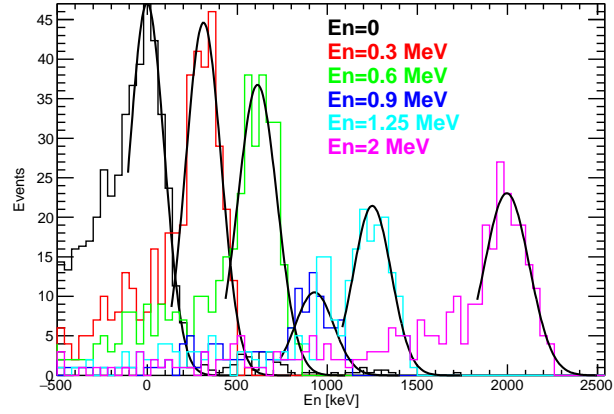}
\caption{\label{fig:ptb_edep_tot_all}Reconstructed neutron energy spectra
for six runs with different neutron reference energy indicated by the labels on the plot.
Each distribution was fitted by a Gaussian shown by black curves to determine
peak position and width. The fit excluded l.h.s. shoulder affected by large
energy loss.}
\end{center}
\end{figure}

\begin{figure}[!h]
\begin{center}
\includegraphics[bb=3.5cm 0cm 20cm 26cm, scale=0.35, angle=270]{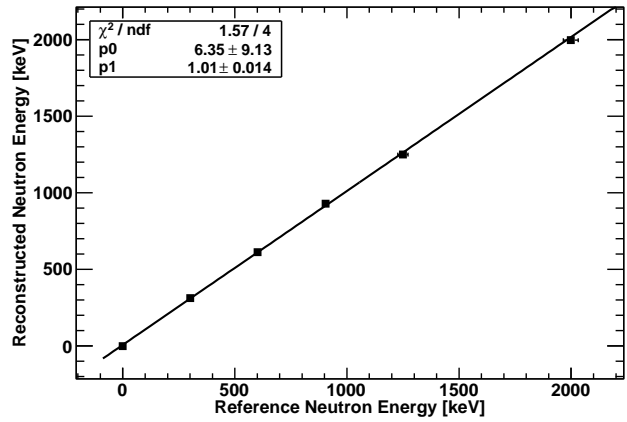}
\caption{\label{fig:ptb_enrec_enref}Correlation between the reference and reconstructed neutron energy
with statistical uncertainty of Gaussian fit and beam energy spread (converted into RMS).
The solid line shows best linear fit whose parameters and $\chi^2$ are
given in the embedded pad.}
\end{center}
\end{figure}

%%% Neutron Energy Reconstruction
Using calibrated deposited energies in two diamonds we reconstructed neutron energy spectra.
To this end we summed up deposited energies in two diamonds, corrected them
for the average total energy loss by both particles (245 keV)
and subtracted conversion reaction $Q$-values (4.784 MeV).
The comparison of neutron spectra measured
in different runs is shown in Fig.~\ref{fig:ptb_edep_tot_all}.
All distributions exhibit a Gaussian-like peak at the incident neutron energy
extended at its l.h.s. by a tail due to excessive energy loss of large angle
$\alpha$ production events. The positions and widths of the peaks were
extracted by a Gaussian fit of the regions unaffected by the energy loss.
The correlation between the reference and reconstructed neutron energy with
their uncertainties is shown in Fig.~\ref{fig:ptb_enrec_enref} along with its linear fit.
The fit gives the correlation coefficient between reference and reconstructed
neutron energy compatible with unity within one standard deviation
and offset of energy scale compatible with zero within one standard deviation.
In absolute scale the deviation of the reconstructed neutron energy
from the reference one does not exceed 11 keV (except for the low statistics 0.9 MeV run where it reaches 24 keV),
the value comparable to the energy spread of the beam given in Table~\ref{tab:beam_target}
and to the statistical uncertainty of Gaussian fit.


\begin{figure}[!h]
\begin{center}
\includegraphics[bb=3.5cm 0cm 20cm 26cm, scale=0.35, angle=270]{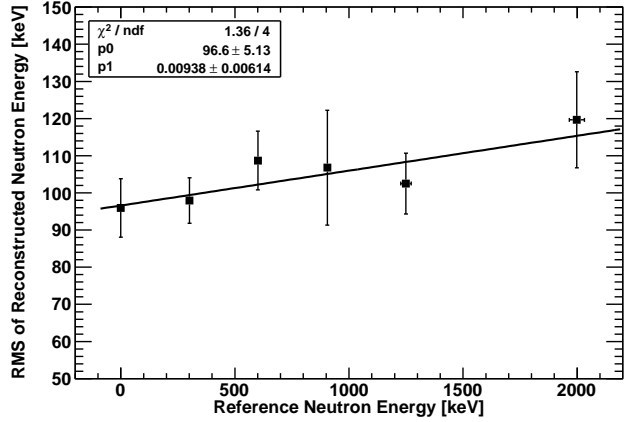}
\caption{\label{fig:ptb_enrec_rms}RMS deviation of the reconstructed neutron energy
obtained from the Gaussian fits in Fig.~\ref{fig:ptb_edep_tot_all}
after subtraction of beam energy spread.}
\end{center}
\end{figure}

RMS deviation of the reconstructed neutron energy is shown in Fig.~\ref{fig:ptb_enrec_rms},
where the contribution of the beam energy spread was subtracted in quadrature
(although it gave almost negligible correction).
The average RMS was about 100 keV made of 50 keV due to electronics resolution
and the remaining 85 keV related to energy loss (expected 30-40 keV).
Slight increase of RMS with neutron energy at the level of one standard deviation (9 keV/MeV)
can be explained, perhaps, by a broader distribution of $\alpha$ energies.

%

%% Geant4
The neutron energy reconstruction described above was mostly data-driven
and required some numerical modeling only to establish a better insight
on the unexpected energy loss tail of $\alpha$s.
For quantitative comparison of absolute event rates however a comprehensive
simulation of the experiment was mandatory.
The response of the spectrometer to these spectra was modeled using Geant version 4.10.3~\cite{geant4}
Monte Carlo simulation software. The spectrometer geometry was carefully described
by simulated volumes, although assuming ideal uniform surfaces of diamond, Au and LiF layers.
The neutrons with the calculated spectra were generated in front of the detector
in plane wave approximation. Event selection procedure was taken from the real data analysis.
An additional ad hoc smearing of the deposited energy was applied to the simulated events
to account for the noise of electronics. This was performed by adding a Gaussian distribution
with zero mean and RMS values measured from the baseline width.

\begin{figure}[!ht]
\begin{center}
\includegraphics[bb=3.5cm 2cm 20.8cm 25cm, scale=0.35, angle=270]{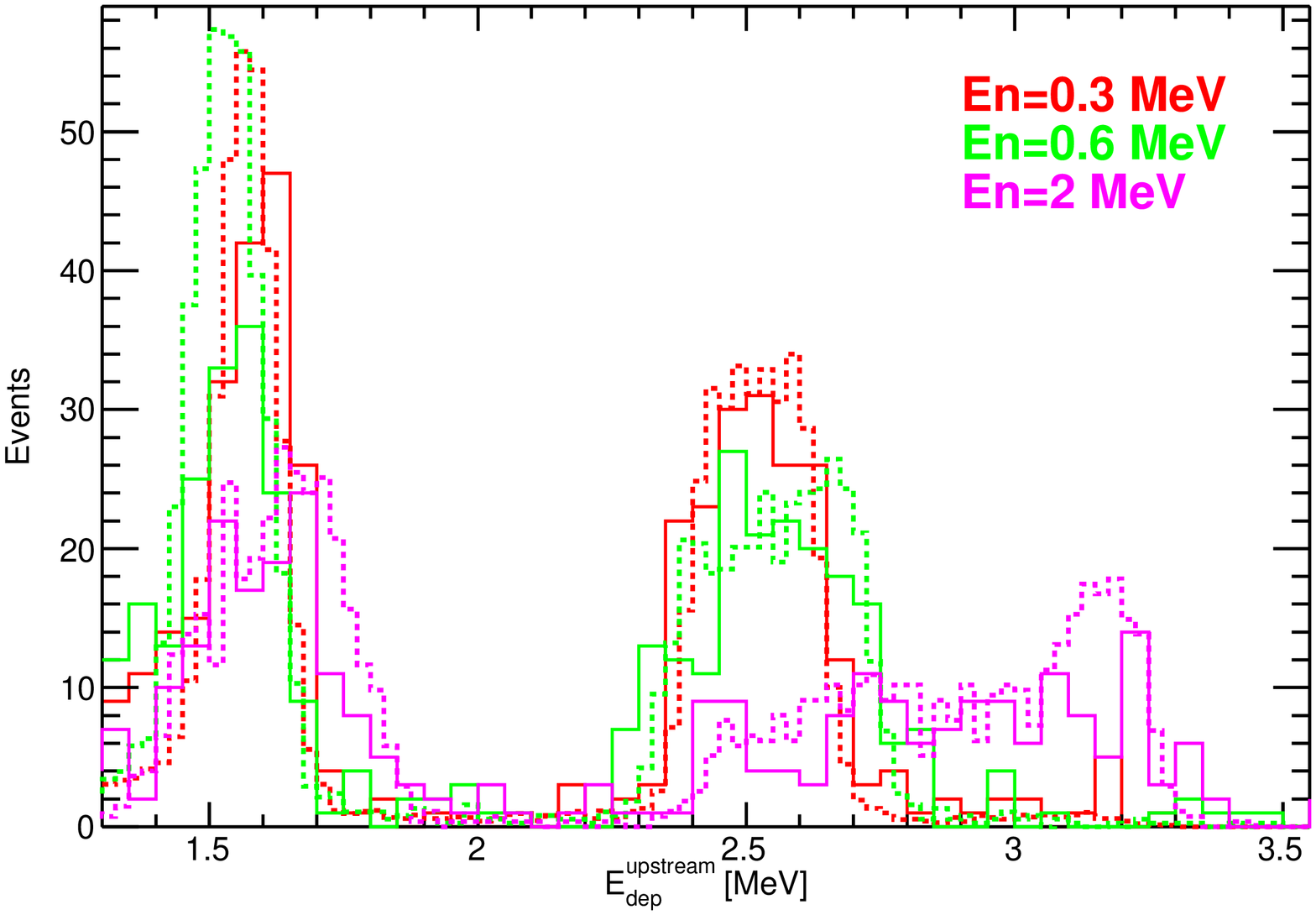}\\
\includegraphics[bb=3.5cm 2cm 20.8cm 25cm, scale=0.35, angle=270]{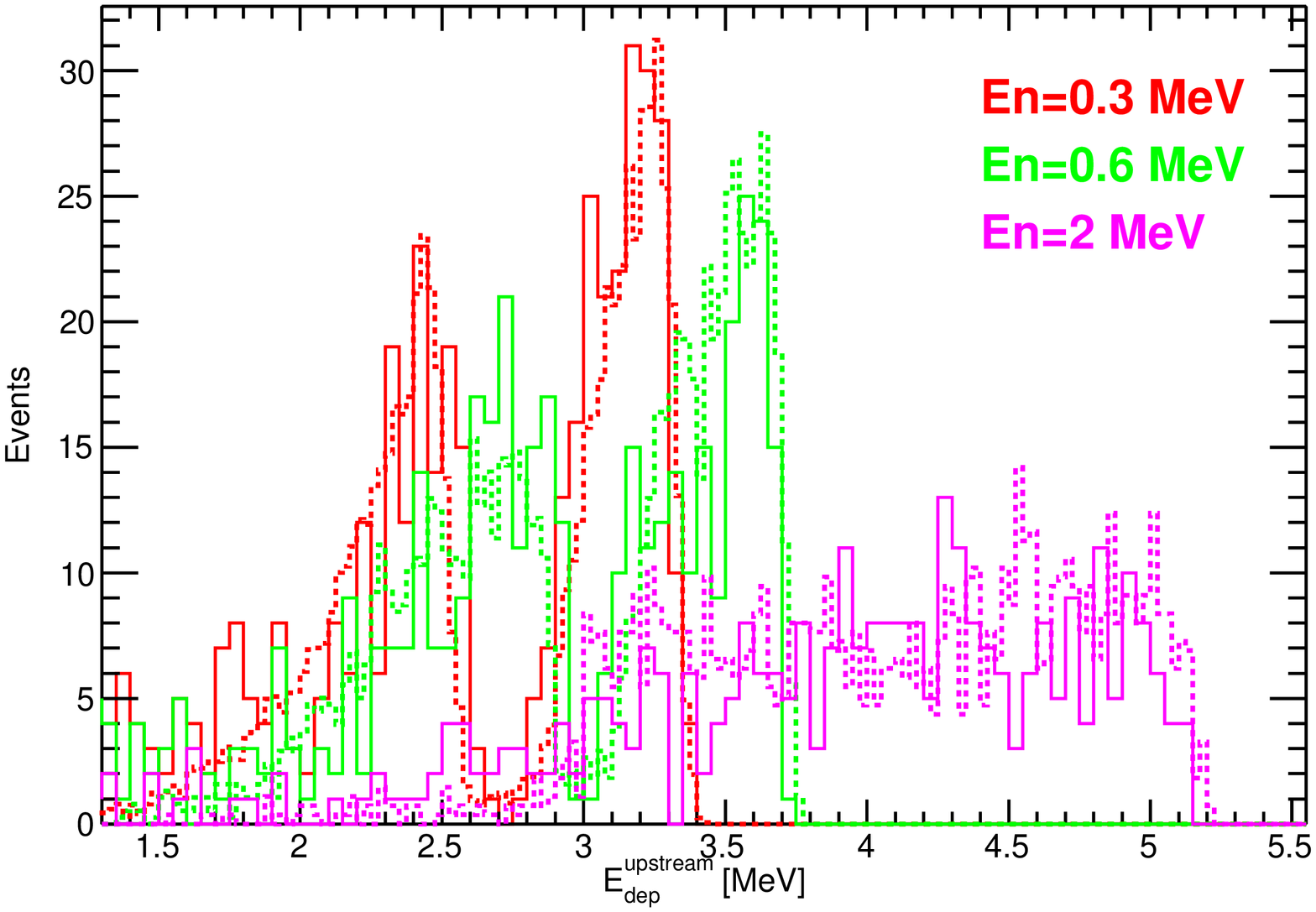}
\caption{\label{fig:ptb_edep12_geant}The same as in Fig.~\ref{fig:ptb_edep12_en}
but in comparison with Geant4 simulations shown by dashed lines.
Geant4 simulations are normalized to the measured neutron fluences and number of generated events.}
\end{center}
\end{figure}
%

%%% Absolute Efficiency
In order to extract the absolute efficiency of the spectrometer we used high precision neutron
fluence values provided by PTB monitoring system. The flux was carefully calibrated for every beam energy,
except for 0.9 MeV where it was calculated from 0.3 MeV calibration, at the beginning of run.
The measured fluence had systematic precision of 2-3\% estimated in Refs.~\cite{Schlegel_3,Schlegel_4}.
Instead, the expected efficiency of the spectrometer was obtained from Geant4 simulations
described above,
as well as from analytical calculations based on $n(^6Li,t)\alpha$ reaction cross section
and known neutron spectra.
The neutron energy distributions, produced by the proton beam impinging on the LiF or TiT targets,
were calculated using a dedicated Monte Carlo code, modelling the specifics of the PTB neutron-producing
targets~\cite{Schlegel_5}. The software simulates both ion and neutron transports
and calculates the neutron energy distribution,
taking into account the ion energy and angular straggling, and the neutron collisions
with the target structural elements. Typically, the produced energy spectrum represents
a narrow peak, centered at the nominal neutron energy, with a width depending mainly
on the target thickness, and a tail, accounting for about 1\%-5\% of the total yield,
that extends to lower energies. This tail component results from neutron scattering
in the target assembly, in particular in the target backing.

\begin{figure}[!ht]
\begin{center}
\includegraphics[bb=3.5cm 0cm 20cm 26cm, scale=0.35, angle=270]{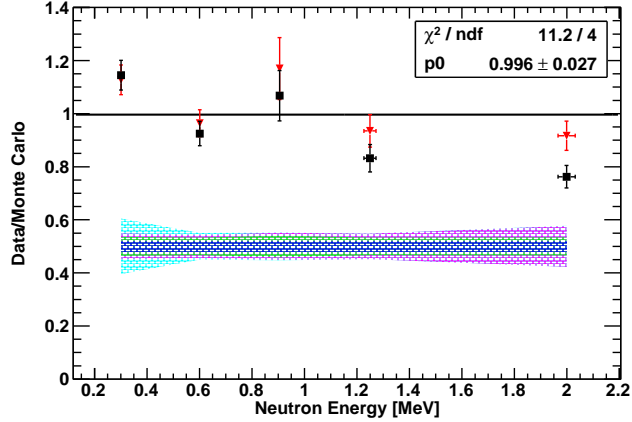}
\caption{\label{fig:ptb_abs_eff}Ratio of the measured and expected efficiencies
of the spectrometer for different neutron energies with its statistical (error bars)
and systematic (hatched bands) uncertainties.
Black points use analytically calculated efficiency, while red points are obtained
from Geant4 simulations.
Blue band represents systematic uncertainty due to the source to detector distance,
green band is due to flux calibration uncertainty,
magenta band is due to ENDF VIII $n(^6Li,t)\alpha$ cross section uncertainty~\cite{endf8},
cyan band is due to absolute neutron energy uncertainty.}
\end{center}
\end{figure}

Geant4 simulations also used these neutron spectra as the input.
The ratio between measured and expected absolute efficiencies is shown in Fig.~\ref{fig:ptb_abs_eff}.
Because of limited statistics of 0.9 MeV measurement the contribution of $^7$Be$^*$(0.43 MeV)
excited state (opening above 0.65 MeV of neutron energy) was subtracted using numerical calculations.
This contribution of 8.4\% was obtained from the $^7$Be$^*$(0.43 MeV) production branching
at zero degrees taken from Ref.~\cite{Liskien} and the ratio of neutron detection
efficiencies at 0.9 and 0.4 MeV.

%

\begin{figure}[!ht]
\begin{center}
\includegraphics[bb=3.5cm 6cm 20cm 24cm, scale=0.35]{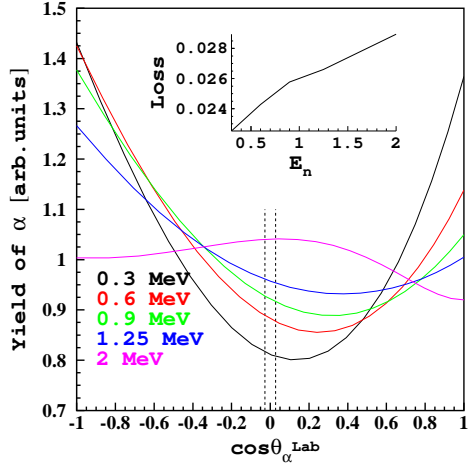}
\caption{\label{fig:li6xs_ang_loss}Angular distribution of $\alpha$ in Lab frame
in $n(^6Li,t)\alpha$ reaction from Ref.~\cite{endf8} for different neutron energies.
The hole in spectrometer acceptance due to the 50 $\mu$m gap is indicated
by the two dashed line. The embedded pad shows the the corresponding efficiency loss
as a function of neutron energy.}
\end{center}
\end{figure}

%%% Efficiency Comparion
The measured  efficiency came out to be compatible with Geant4 based expectations
within statistical uncertainties for all five neutron energies.
Analytical calculations were sufficiently accurate up to 1 MeV,
above this energy they systematically underestimated the efficiency.
In the calculations we neglected the acceptance deficiency due to
the 50 $\mu$m gap between two diamonds. Instead, Geant4 simulation
had this gap implemented along with non-trivial angular distribution $n(^6Li,t)\alpha$
reaction products from Ref.~\cite{endf8}. The change of angular distribution with neutron energy,
shown in Fig.~\ref{fig:li6xs_ang_loss}, indeed resulted in increasing efficiency
loss, however the magnitude of this effect was lower than 0.5\%.
Instead the Lorentz boost effectively reduced the angle between $t$ and $\alpha$
(produced back-to-back in thermal neutron conversion) allowing both particle
to go forward, violating the coincidence trigger condition.
To estimate order of magnitude of this effect we evaluated analytically the maximum
angle reduction to be 9.5, 13.4, 16, 18.5 and 22.3 degrees 
for 0.3, 0.6, 0.9, 1.25 and 2 MeV neutron energy, respectively.
% Delta cos CM: 0.095, 0.13, 0.155, 0.18 and 0.215
These values, after subtraction of 50 $\mu$m gap and multiplication by the number of final state particles (two),
provide right estimate of the observed inefficiency.
It is worth noting that this inefficiency due to Lorentz boost is maximized
for neutrons impacting the detector along its normal, like in the present experiment,
while for an isotropic irradiation, e.g. in reactor core, it will be considerably reduced.

Major systematic uncertainties on the efficiency ratio came from neutron flux calibrations,
the source to spectrometer distance uncertainty,
uncertainty of $n(^6Li,t)\alpha$ cross section and for 0.3 MeV measurement
from the neutron energy uncertainty.
$\gamma$ background was very well separated from the neutron spectra
by high $Q$-value of conversion reaction as one can see in Fig.~\ref{fig:ptb_edep_tot_th},
and event rates were low compared to 64 ns coincidence window to give any accidental.
Because of few MeV wide integration interval of the reconstructed neutron
energy peak small uncertainties ($<$11 keV) on the energy calibrations were negligible.
Neutron beam energy, except for the natural width of the spectra described above,
had an uncertainty of $<4$ keV due to proton beam energy and target thickness.
This resulted in 5.6\% systematic uncertainty on the $n(^6Li,t)\alpha$ cross section
at 0.3 MeV, while negligible at higher energies.
The combination of all these uncertainties in quadrature gave the overall
uncertainties varying from 3 to 6\% with the average of 4\%.

%%%%%%%%%CONCLUSIONS%%%%%%%%%%%%%%%%%
\section{Conclusions}\label{sec:conclusions}
Recently developed prototype of the compact neutron spectrometer was calibrated
at PTB ISO certified neutron source.
The spectrometer is based on $^6$Li converter and a sandwich of
two diamond detectors measuring both products of neutron conversion
in coincidence. The spectrometer has relatively low detection efficiency
and aimed mostly for the high flux environments like those in research reactors.
The calibrated prototype was based on 100 $\mu$m thick 4$\times$4 mm$^2$
crystals produced by Applied Diamond Inc~\cite{applied_diamond}.
Diamond sensors had thin ohmic Au contacts reducing energy loss by
$t$ and $\alpha$ products of neutron conversion on $^6$Li
and suppressing space charge effects.

Measurements were performed at five different neutron energies
varying from 0.3 up to 2 MeV.
The reconstructed neutron energy agrees with the reference values
within statistical uncertainties or better than 11 keV in absolute value.
The measured detection efficiency agrees with expectations
within statistical and systematic uncertainties, providing
critical test of the developed spectrometer to the overall precision of
3\% statistical and 4\% systematic.
The expectations were based on $n(^6Li,t)\alpha$ cross section
obtained from ENDF database~\cite{endf8}
and Geant4 simulations, important only at energies above 1 MeV.

%%%%%%%ACKNOWLEDGEMENTS%%%%%%%%%%%%%%%%%%%%%%%%%%%%%%%%%%%%%%%%%%%%%%%
\section*{Acknowledgements}
Authors would like to acknowledge the excellent support provided during the experiment
by the staff and technical services of PTB facility.
This experiment has been supported by Istituto Nazionale di Fisica Nucleare
through the strategic project ``INFN Energy'',
by Centro Fermi - Museo Storico della Fisica e Centro Studi e Ricerche ``Enrico Fermi''
and by the European Atomic Energy Community's (Euratom) Seventh Framework Program 2007-2013
under Transnational Access workpackage of the Project CHANDA (Grant Agreement no. 605203).
Detailed Geant4 simulations were performed on the HPC cluster ``OCAPIE''
funded by Compagnia di San Paolo Foundation.
%%%%%%%%%%%%%%%%%%%%%%%%%%%%%%%%%%%%%%%%%%%
\bibliographystyle{elsarticle-num}
\bibliography{ptb_calib}

\begin{thebibliography}{10}
\expandafter\ifx\csname url\endcsname\relax
  \def\url#1{\texttt{#1}}\fi
\expandafter\ifx\csname urlprefix\endcsname\relax\def\urlprefix{URL }\fi
\expandafter\ifx\csname href\endcsname\relax
  \def\href#1#2{#2} \def\path#1{#1}\fi

\bibitem{cvd_rad_hard}
W.~de~Boer, et~al., Phys. Status Solidi 204 (2007) 3009.

\bibitem{sdw_calib}
M.~Osipenko, et~al., Nucl. Instr. and Meth. A 799 (2015) 207.

\bibitem{sdw_tapiro}
M.~Osipenko, et~al., Test of a prototype neutron spectrometer based on diamond
  detectors in a fast reactor, in: Proc. 4th International Conference on
  Advancements in Nuclear Instrumentation Measurement Methods and their
  Applications (ANIMMA 2015), IEEE Nucl.Sci.Symp.Conf.Rec., Lisbon, Portugal,
  2015.
\newblock \href {http://dx.doi.org/10.1109/ANIMMA.2015.7465605}
  {\path{doi:10.1109/ANIMMA.2015.7465605}}.

\bibitem{sdw_dt}
M.~Osipenko, et~al., Nucl. Instr. and Meth. A 817 (2016) 19.

\bibitem{sdw_upg}
M.~Osipenko, et~al., Nucl. Instr. and Meth. A 883 (2018) 14.

\bibitem{applied_diamond}
\href{http://usapplieddiamond.com}{{Applied Diamond Inc}}.
\newline\urlprefix\url{http://usapplieddiamond.com}

\bibitem{Trucchi1}
D.~Trucchi, et~al., IEEE Electron Device Lett. 33 (2012) 615.

\bibitem{Trucchi2}
M.~Girolami, et~al., Phys. Status Solidi A 212 (2015) 2424.

\bibitem{Trucchi3}
M.~Rebai, et~al., Journal of Instrumentation 10 (2015) {C03016}.

\bibitem{Trucchi4}
C.~Cazzaniga, et~al., Nucl. Instr. and Meth. B 405 (2017) 1.

\bibitem{ptb_n}
\href{https://www.ptb.de/cms/en/ptb/fachabteilungen/abt6/fb-64.html}{{Physikal%
isch-Technische Bundesanstalt (PTB), Neutron Metrology}}.
\newline\urlprefix\url{https://www.ptb.de/cms/en/ptb/fachabteilungen/abt6/fb-6%
4.html}

\bibitem{Nolte_1}
R.~Nolte, D.~J. Thomas, Metrologia 48 (2011) S274--S291.

\bibitem{Roberts_2}
N.~Roberts, et~al., Radiation Measurements 45 (2010) 1151--1153.

\bibitem{Schlegel_3}
D.~Schlegel, S.~Guldbakke, Neutron fluence measurements with a recoil
  proportional counter, Tech. Rep. {PTB-6.41-2002-03}, {PTB} (2002).

\bibitem{Schlegel_4}
D.~Schlegel, S.~Guldbakke, Neutron fluence measurements with a recoil proton
  telescope, Tech. Rep. {PTB-6.41-2002-04}, {PTB} (2002).

\bibitem{geant4}
S.~Agostinelli, et~al., Nucl. Instr. and Meth. 506 (2003) 250.

\bibitem{Schlegel_5}
D.~Schlegel, Target users manual, Tech. Rep. {PTB-6.42-05-2}, {PTB} (2005).

\bibitem{Liskien}
H.~Liskien, A.~Paulsen, Atomic Data and Nuclear Data Tables 15 (1975) 57.

\bibitem{endf8}
M.~Chadwick, et~al., Nucl. Data Sheets 112 (2011) 2887.

\end{thebibliography}

\end{document}